\documentclass[a4,12pt]{article}
\newcommand{\nn}{\nonumber\\}
\newcommand{\bra}[1]{\left<#1 \right|}
\newcommand{\ket}[1]{\left| #1 \right>}
\newcommand{\abs}[1]{\left| #1 \right|}

\newcommand{\Q}{Q_{\rm B}}

\renewcommand{\thepage}{}
\makeatletter
\@addtoreset{equation}{section}
\renewcommand{\theequation}{\thesection.\@arabic\c@equation}
\makeatother

\begin{document}
\begin{titlepage}
\title{
\vspace*{-4ex}
\hfill{\normalsize hep-th/0409249}\\
\vspace{4ex}
\bf Marginal Deformations and Closed String Couplings
in Open String Field Theory\\
\vspace{5ex}}
\author{Fumie {\sc Katsumata}\footnote{E-mail address:
 katsumata@asuka.phys.nara-wu.ac.jp},\ 
Tomohiko {\sc Takahashi}\footnote{E-mail address:
 tomo@asuka.phys.nara-wu.ac.jp}\ \ \ 
and \ Syoji {\sc Zeze}\footnote{E-mail address:
zeze@asuka.phys.nara-wu.ac.jp}
\vspace{3ex}\\
{\it Department of Physics, Nara Women's University, Japan}}
\date{September, 2004}
\maketitle
\vspace{7ex}

\begin{abstract}
\normalsize
\baselineskip=19pt plus 0.2pt minus 0.1pt
We investigate analytic classical solutions in open string field
theory which are constructed in terms of marginal operators.
In the classical background, we
evaluate a coupling between an on-shell closed string state and the open
string field. The resulting coupling
exhibits periodic behavior as expected from the marginal boundary
deformation of background Wilson lines or a marginal tachyon lump.
We confirm that the solutions in open string field
theory correspond to a class of marginal deformations in conformal
field theory. 
\end{abstract}
\end{titlepage}

\renewcommand{\thepage}{\arabic{page}}
\setcounter{page}{1}
\baselineskip=19pt plus 0.2pt minus 0.1pt

\section{Introduction}

String field theory possesses enormously large gauge structure which 
can never be found in low energy effective field theories. 
String field, its integration and the
BRS charge have similar properties to a connection,
the integration of differential forms and
the exterior derivative in ordinary differential geometry. The action of
open string field theory is analogue to 
the integration of the Chern-Simons three form, and then the theory is
invariant under analogous gauge symmetry
\cite{rf:CSFT,rf:SCSFT}. Moreover, the gauge 
symmetry becomes much larger than that of
effective theories due to infinite component fields of string field and
their product with complicated non-local structure.

The equation of motion in open string field theory
means that its solutions can be expressed
as an analogue to a flat connection in differential geometry 
\cite{rf:CSFT}.
By analogy with field theory, we expect
that classical solutions can be classified by topologically inequivalent 
gauge group elements in string field theory.
For example, the tachyon vacuum solution should be
obtained as a kind of large gauge transformations of the perturbative
vacuum as proposed in ref.~\cite{rf:TT2}.
Hence, gauge structure of string field theory may be an important and
fundamental concept for the classification of non-perturbative vacua in
string field theory.

It was proposed in refs.~\cite{rf:TT2} and \cite{rf:TT1} that
an analytic classical solution is represented as a locally pure gauge
configuration in open string field 
theory, and it can be related to a marginal boundary deformation in
conformal field theory. This correspondence between gauge
transformations and marginal deformations is a natural generalization of
that of effective theories. On the other hand, using the level
truncation analysis \cite{rf:SZmarginal}, we can construct a one
parameter family of classical solutions associated with the marginal
operator. Naturally, we expect that these are physically
equivalent solutions and that the solution in the level truncated theory
may be 
expressed also as locally pure gauge form.
However, it is difficult to compare these solutions directly and to
assure their equivalence because of their gauge difference.
In addition, it is impossible to rewrite the
level truncated solution as a pure gauge form because the level
truncation breaks gauge symmetry of string field theory.

Though the level truncated theory is not helpful to find the gauge
structure, the theory is very useful to evaluate vacuum energy for some
classical solutions \cite{rf:KS-tachyon,rf:SZ-tachyon,rf:MT,rf:GR}.
Actually, we find that the vacuum energy of the
marginal solution approaches to zero for all allowed value of the
massless field as the 
truncation level is increased \cite{rf:SZmarginal}. This result is
consistent with the fact that the marginal solution corresponds to a
boundary marginal deformation in conformal field theory. 
In contrast to the truncated case, 
it turns out that the vacuum energy of the analytic
solution is a kind of indefinite quantities \cite{rf:tomo}.
Thus, each one of the solutions has both merits and demerits.

In order to understand both gauge group and vacuum structures in 
string field theory, we should clarify further the
relation of the analytic and truncated solutions, as a first step, 
for marginal deformations. 
For the level truncated solution, it has recently become possible to
construct the energy momentum tensor associated with the marginal
solution and to compare this with the result in conformal field
theory, namely a disk amplitude with a graviton vertex insertion in the
marginal deformed background \cite{rf:SenEM}. 
If these solutions are gauge equivalent, we may provide similar
results for the analytic classical solution. 

In this paper, we will calculate open-closed
string couplings around the vacuum corresponding to the analytic marginal
solutions. We use gauge invariant operators proposed in
ref.~\cite{rf:Zwiebach} to represent the open-closed string coupling
which is related
to a disk amplitude with an additional open string vertex insertion.
As a result,
we reproduce their expected periodic behavior for marginal parameters,
for example creation and annihilation of an
array of D-branes \cite{rf:CKLMPT,rf:RS,rf:Sen}. These results confirm
that the analytic 
marginal solutions are associated with marginal boundary deformations of
the conformal field theory.

In section 2 we will consider a class of analytic solutions
corresponding to Wilson 
lines background and marginal tachyon lumps as in
refs.~\cite{rf:TT2,rf:TT1}.
In section 3 we will derive oscillator expression of the the on-shell closed
string coupling to the open string field. Using this expression, we will
evaluate open-closed string couplings in several classical
backgrounds. 

\section{Classical Solutions and Marginal Deformations}

The action in open string field theory is given by \cite{rf:CSFT}
\begin{eqnarray}
 S[\Psi]=-\frac{1}{g^2}\int \left(\frac{1}{2}\Psi*\Q\Psi
+\frac{1}{3}\Psi*\Psi*\Psi\right). 
\label{Eq:action}
\end{eqnarray}
Under the gauge transformation $\Psi=U*\Q\,U^{-1}+U*\Psi'*U^{-1}$,
the action is transformed as
$S[\Psi]=S[U*\Q\,U^{-1}]+S[\Psi']$.
The functional $U$ is an element of the gauge group in which the
multiplication law is given by the star product. 
We expect that the action is invariant under gauge
transformations implemented
by the gauge functional which is homotopically equivalent to the
identity string field $I$. Because such gauge functionals can be
generated by 
integrating infinitesimal gauge transformations from the 
identity, and the action is invariant under the infinitesimal
gauge transformation
$\delta\Psi=\Q\Lambda+\Psi*\Lambda-\Lambda*\Psi$ \cite{rf:GJ,rf:IOS,
rf:LPP}. From variational principle, the equation of motion becomes
$\Q\Psi+\Psi*\Psi=0$.  

We consider the case of no Chan-Paton degrees of freedom. We
single out a direction and write its string coordinate as
$X(z,\bar{z})=(X(z)+X(\bar{z}))/2$.
Then we can construct a classical
solution using the operator $\partial X(z)$ \cite{rf:TT2,rf:TT1}:
\begin{eqnarray}
\label{Eq:sol}
 \Psi_0(\lambda)=-\lambda V_L(F)I+\frac{1}{4}\lambda^2 C_L(F^2)I,
\end{eqnarray}
where $\lambda$ is a real parameter and 
the operators $V_L(F)$ and $C_L(F^2)$ are defined as
\begin{eqnarray}
\label{Eq:VLCL}
 V_L(F)=\int_{C_{\rm left}}\frac{dz}{2\pi i}F(z)\,
\frac{i}{2\sqrt{\alpha'}} c(z)\partial X(z),\ \ \ 
 C_L(F^2)=\int_{C_{\rm left}}\frac{dz}{2\pi i}F(z)^2\,c(z).
\end{eqnarray}
Here $C_{\rm left}$ denotes the integration path along the left half of
a string and $F(z)$ is a function satisfying $F(-1/z)=z^2F(z)$ and
$F(\pm i)=0$.\footnote{We denote the complex coordinate $w$ in
refs.~\cite{rf:TT2,rf:TT1} by $z$.}
The function $F(z)$ corresponds to
$F_+^{(1)}(z)$ in ref.~\cite{rf:TT2} and we will see later that
$F_+^{(1)}(z)$ provides a solution with the same physical property as
$F_-^{(1)}(z)$. 
\footnote{We can show that $\Psi_0(\lambda)$ obeys the equation of
motion by using the commutation relation
\begin{eqnarray}
 \{V_L(F),\,V_L(F)\}=-\frac{1}{2}\{\Q,\,C_L(F^2)\},
\end{eqnarray}
and the following properties
\begin{eqnarray}
&& \left(V_R(F)A\right)*B=-(-1)^{\abs{A}}A*V_L(F)B,\nn
&& \left(C_R(F)A\right)*B=-(-1)^{\abs{A}}A*C_L(F)B,\\
&& V_R(F)I=-V_L(F)I,\ \ \ C_R(F)I=-C_L(F)I.
\end{eqnarray}
}

We introduce an operator using $X(z)$ and $F(z)$: 
\begin{eqnarray}
\label{Eq:XL}
 X_L(F) =\int_{C_{\rm left}}\frac{dz}{2\pi i}F(z)X(z).
\end{eqnarray}
Using properties of $X_L(F)$ in ref.~\cite{rf:TT2},
we can rewrite locally the solution as a pure gauge form:
\begin{eqnarray}
 \Psi_0(\lambda)=U(\lambda)*\Q\,U(\lambda)^{-1},
\end{eqnarray}
where the group element $U(\lambda)$ is
\begin{eqnarray}
\label{Eq:Ulamd}
U(\lambda)=\exp\left(
   \frac{i\,\lambda}{2\sqrt{\alpha'}}\,
   X_L(F)\,I\right),
\end{eqnarray}
and we have defined as $\exp A=I+A+A*A/2!+\cdots$.
This is a local expression because the operator $X_L(F)$ includes the
zero mode $\hat{x}$ of the string coordinate.

If we expand the string field as $\Psi=\Psi_0(\lambda)+\tilde{\Psi}$,
the action becomes
\begin{eqnarray}
\label{Eq:shiftaction}
 S[\Psi]=S[\Psi_0(\lambda)]
-\frac{1}{g^2}\int \left(\frac{1}{2}\tilde{\Psi}*
\Q'(\lambda) \tilde{\Psi}
+\frac{1}{3}\tilde{\Psi}*\tilde{\Psi}*\tilde{\Psi}\right).
\end{eqnarray}
The modified BRS charge $\Q'(\lambda)$ is given by
\begin{eqnarray}
\label{Eq:newBRS}
 \Q'(\lambda)=\Q-\lambda\left(V_L(F)+V_R(F)\right)
+\frac{\lambda^2}{4}\left(C_L(F^2)+C_R(F^2)\right),
\end{eqnarray}
where the operators $V_R(F)$ and $C_R(F^2)$ are defined
by replacing $C_{\rm left}$  in (\ref{Eq:VLCL}) by  $C_{\rm right}$
which is the path along the right half of a string.
As discussed in ref.~\cite{rf:TT2}, the action for the quantum
fluctuation is transformed to the original form by the string field
redefinition:
\begin{eqnarray}
\label{Eq:redef}
 &&
\tilde{\Psi}=e^{B(\lambda)}\Psi',\nn
&&B(\lambda)=
\frac{i \lambda}{2 \sqrt{\alpha'}}X_L(F)
+\frac{i \lambda}{2 \sqrt{\alpha'}}X_R(F),
\end{eqnarray}
where the operator $X_R(F)$ is defined
by replacing $C_{\rm left}$ in (\ref{Eq:XL}) by  $C_{\rm right}$. For
this redefinition, one of important equations is
$\Q=e^{-B(\lambda)}\Q'(\lambda)e^{B(\lambda)}$. 
Thus, the action $S[\Psi]$ is transformed to the original form
$S[\Psi']$ with the constant term $S[\Psi_0(\lambda)]$. This result
suggests that 
the string field expansion and the redefinition can be
regarded as a gauge transformation in string field theory. Indeed, the
expansion and redefinition of the string field can be written locally
in terms of the gauge group element (\ref{Eq:Ulamd}):
\begin{eqnarray}
 \Psi&=&\Psi_0(\lambda)+e^{B(\lambda)}\Psi'\nn
     &=&U(\lambda)*\Q\,U(\lambda)^{-1}
+U(\lambda)*\Psi'*\,U(\lambda)^{-1}.
\end{eqnarray}

To exhibit the physical meaning of the classical solution, we choose 
the function $F(z)=(z+1/z)/z$ as the simplest case.
Though there is other possibility for
$F(z)$, we can transform the solution for other functions to this
simplest case by string field redefinitions or gauge transformations.
In this case, the solution (\ref{Eq:sol})
has a well-defined Fock space expression
\cite{rf:TT2,rf:TT1}:
\begin{eqnarray}
\label{Eq:sftcond}
 \ket{\Psi_0(\lambda)}=\frac{\lambda^2}{2\pi}c_1\ket{0}
  -\frac{\lambda}{\sqrt{2}\pi}c_1\alpha_{-1}\ket{0}
  +\frac{\lambda^2}{6\pi}c_{-1}\ket{0}
  +\frac{\lambda^2}{2\pi}c_1L_{-2}\ket{0}+\cdots,
\end{eqnarray}
and the operator $B(\lambda)$ can be written
\begin{eqnarray}
\label{Eq:Blamd}
B(\lambda)=-\frac{\lambda}{\sqrt{2}}(\alpha_1-\alpha_{-1}).
\end{eqnarray}
Here, $\alpha_n$ and $L_n$ are oscillators of $X(z)$ and total Virasoro
generators, respectively, and $\ket{0}$ is the $SL(2,R)$ invariant
vacuum and the abbreviation denotes higher level states.
Using oscillator representation, we write the string field as
\begin{eqnarray}
 \ket{\Psi}=\phi(x)\,c_1\ket{0}+A(x)\,c_1\alpha_{-1}\ket{0}
+A_i(x)\,c_1\alpha_{-1}^i\ket{0}+\cdots,
\end{eqnarray}
where $\alpha_n^i\ (i=0,\cdots,24)$ is oscillators of other
directions except $X(z)$ and the abbreviation denotes higher level
component fields. 
Substituting (\ref{Eq:Blamd}) into (\ref{Eq:redef}),
we can represent the string field redefinition
using component fields:
\begin{eqnarray}
 \tilde{\phi}(x)&=&e^{-\frac{\lambda^2}{4}}\,\phi'(x)
-\frac{\lambda}{\sqrt{2}}e^{-\frac{\lambda^2}{4}}A'(x)+\cdots, \\
 \tilde{A}(x)&=&\frac{\lambda}{\sqrt{2}}
       e^{-\frac{\lambda^2}{4}}\,\phi'(x)
+\left(1-\frac{\lambda^2}{2}\right)
      e^{-\frac{\lambda^2}{4}}A'(x)+\cdots, \\
 \tilde{A_i}(x)&=&e^{-\frac{\lambda^2}{4}}A_i'(x)+\cdots.
\end{eqnarray}
It should be noticed that, in these equations, all coefficients of
component fields have regular expression for any
$\lambda$,\footnote{This result follows from the fact that a
normal ordered form of the operator $\exp B(\lambda)$ is well-defined
regular expression for any $\lambda$.}
and then the string field redefinition (\ref{Eq:redef})
is well-defined for all $\lambda$. 

We expect that this solution corresponds to a
the marginal deformation of Wilson lines because $\Psi_0(\lambda)$
contains a vacuum expectation value of a massless vector state,
$c_1\alpha_{-1}\ket{0}$, in the expression (\ref{Eq:sftcond}). 
In fact, if we introduce Chan-Paton indices as
$\Psi_{ij}$, the string field redefinition from $\tilde{\Psi}$ to
$\Psi'$ is written as
\begin{eqnarray}
&&
 \tilde{\Psi}_{ij}=e^{B(\lambda_i,\lambda_j)}\Psi'_{ij},
\end{eqnarray}
where the operator $B(\lambda_i,\lambda_j)$ is given by
\begin{eqnarray}
B(\lambda_i,\lambda_j)&=&
-\frac{i \lambda_i}{2 \sqrt{\alpha'}}X_L(F)
-\frac{i \lambda_j}{2 \sqrt{\alpha'}}X_R(F),\nn
&=& i\,\frac{\lambda_i-\lambda_j}{\pi\sqrt{\alpha'}}\hat{x}
-\frac{\lambda_i+\lambda_j}{2\sqrt{2}}(\alpha_1-\alpha_{-1})+\cdots.
\end{eqnarray}
Hence, this string field redefinition changes the momentum of
$\tilde{\Psi}_{ij}$ as $p\rightarrow p+(\lambda_i-\lambda_j)/\pi
\sqrt{\alpha'}$ and this is the same effect as background Wilson
lines \cite{rf:TT2,rf:TT1}. 

Since the analytic solution $\Psi_0(\lambda)$ corresponds to the
marginal deformation, the vacuum energy $S[\Psi_0(\lambda)]$ 
should be zero as expected from that of the level truncated solution in
ref.~\cite{rf:SZmarginal}. Formally, we can evaluate the vacuum energy
in the following \cite{rf:KZ}. We differentiate $S[\Psi_0(\lambda)]$
with respect to $\lambda$:
\begin{eqnarray}
\label{Eq:dS}
 \frac{d}{d\lambda}S[\Psi_0(\lambda)]=
-\frac{1}{g^2}\int\frac{d\Psi_0(\lambda)}{d\lambda}*\left(
\Q\Psi_0(\lambda)+\Psi_0(\lambda)*\Psi_0(\lambda)\right)=0,
\end{eqnarray}
where we have used the fact that $\Psi_0(\lambda)$ obeys the equation of
motion. Then, we find that $S[\Psi_0(\lambda)]$ is a constant and
independent of $\lambda$. Since $\Psi_0(0)=0$ and then $S[\Psi_0(0)]=0$,
we verify that the vacuum energy becomes zero, that is
$S[\Psi_0(\lambda)]=0$. However, if we evaluate $S[\Psi_0(\lambda)]$
directly in terms of its oscillator representation, we find that the
vacuum energy is an indefinite quantity given by zero from the ghost
sector times infinity from the matter sector  as discussed in
ref.~\cite{rf:tomo}. 
Then, there should be exist a kind of regularization method to calculate
the vacuum energy directly. Though it is an open question to find the
regularization, we will formally handle similar indefinite quantities in
the next section.

We can construct other analytic classical solutions corresponding to
different marginal deformations.
As an example, we consider the
case that the direction $X$ is compactified on a circle of critical
radius. At the critical radius, there are the following three conserved
currents 
\begin{eqnarray}
 J^1(z)&=& \frac{i}{2\sqrt{\alpha'}}\partial X(z),\\
 J^2(z)&=& \frac{1}{2}\left(
  e^{\frac{i}{\sqrt{\alpha'}}X(z)} +e^{-\frac{i}{\sqrt{\alpha'}}X(z)}
 \right),\\
 J^3(z)&=& \frac{1}{2i}\left(
  e^{\frac{i}{\sqrt{\alpha'}}X(z)} -e^{-\frac{i}{\sqrt{\alpha'}}X(z)}
 \right),
\end{eqnarray}
and these currents generate a $SU(2)$ symmetry in string theory.
If we expand the currents 
as\footnote{Using oscillators of $X(z)$, the current $J^1(z)$
is expanded as $J^1(z)=\sum_n \alpha_n z^{-n-1}/\sqrt{2}$.}
\begin{eqnarray}
 J^2(z)=\frac{1}{\sqrt{2}}
    \sum_{n=-\infty}^\infty \beta_n z^{-n-1},\ \ \ 
 J^3(z)=\frac{1}{\sqrt{2}}
    \sum_{n=-\infty}^\infty \gamma_n z^{-n-1},
\end{eqnarray}
these oscillators satisfy the commutation relations 
\begin{eqnarray}
\label{Eq:comrel}
&&
[\alpha_m,\,\alpha_n]=m\delta_{m+n,0},\ \ \ 
[\beta_m,\,\beta_n]=m\delta_{m+n,0},\ \ \ 
[\gamma_m,\,\gamma_n]=m\delta_{m+n,0},\ \ \ \nn
&&
 \left[\alpha_m,\,\beta_n\right]=i\,\gamma_{m+n},\ \ \ 
 \left[\beta_m,\,\gamma_n\right]=i\,\alpha_{m+n},\ \ \ 
 \left[\gamma_m,\,\alpha_n\right]=i\,\beta_{m+n}.
\end{eqnarray}
It can be easily seen that $\alpha_0$, $\beta_0$ and $\gamma_0$ commute
with the BRS charge and these zero modes are conserved
on the three string vertex in the action.
Then, the action (\ref{Eq:action}) is invariant
under the $SU(2)$ rotation
\begin{eqnarray}
 \Psi=\exp\left(i \theta_1 \alpha_0
+i \theta_2 \beta_0 +i \theta_3 \gamma_0 \right)\,\Psi',
\end{eqnarray}
where $\theta_k$ are real parameters.
Acting a $SU(2)$ rotation on (\ref{Eq:sol}),
we can find as a analytic solution,
\begin{eqnarray}
\label{Eq:puregaugek}
 \Psi^k_0(\lambda)=-\lambda V_L^k(F)I+\frac{\lambda^2}{4}C_L(F^2)I,
\end{eqnarray}
where $V_L^k(F)$ are defined by
\begin{eqnarray}
 V^k_L(F)=\int_{C_{\rm left}}\frac{dz}{2\pi i}F(z)\,c(z)\,J^k(z)
\ \ \ (k=1,2,3).
\end{eqnarray}
The $k=1$ case corresponds to the previous marginal configuration.
If we choose $F(z)=(z+1/z)/z$ and we expand the string field as
$\Psi=\Psi_0^2(\lambda)+\tilde{\Psi}$, the action of the 
fluctuation goes back to the original action
by the string field redefinition
\begin{eqnarray}
\label{Eq:Blamdcr} \tilde{\Psi}=e^{B'(\lambda)}\Psi',\ \ \ 
 B'(\lambda)=-\frac{\lambda}{\sqrt{2}}(\beta_1-\beta_{-1}).
\end{eqnarray}
Instead of use of the $SU(2)$ rotation,
we can show more explicitly
that the solution (\ref{Eq:puregaugek})
obeys the equation of motion as discussed in
refs.~\cite{rf:TT2,rf:TT1}.

\section{Closed String Couplings and Marginal Deformations}

In this section we will consider open-closed string couplings 
at the classical vacuum corresponding to the analytic marginal
solutions. 
As in refs.~\cite{rf:Zwiebach,rf:HI,rf:TZ2,rf:AG,rf:GM}, we can
incorporate closed strings by introducing the term
\begin{eqnarray}
\label{Eq:closedterm}
 \bra{V}\ket{\Psi},\ \ \ 
 \bra{V}=\bra{I}V\left(\frac{\pi}{2}\right),
\end{eqnarray}
where $V(\sigma)$ corresponds to an on-shell closed string vertex
operator
$V(\sigma)=c_+(\sigma)c_-(\sigma)O(\sigma)$.
We consider the case that the direction $X$ is compactified on a circle
of radius $R$ and the Neumann boundary condition is imposed on $X$.
If the vertex operator has no derivative operator of this
direction, namely $\partial X, \partial^2 X,\cdots$,
the closed string state with the momentum $m/R$ and the
winding number $w$ is given by
\begin{eqnarray}
\label{Eq:VmwR}
 \bra{V(m,w;R)}
=\bra{V_{c=25}}\otimes \bra{m,w;R} \otimes \bra{V_{gh}},
\end{eqnarray}
where $\bra{V_{c=25}}$ and $\bra{V_{gh}}$ correspond to vertex operators
of the rest of the $c=25$ CFT and the ghost CFT. The state $\bra{m,w;R}$
is given by the definition which refers to the $c=1$ CFT of the
compactified direction:
\begin{eqnarray}
\label{Eq:mwRdef}
 \bra{m,w;R}\ket{\phi}=\Big\langle\,e^{ik_L X(i)+ik_R X(-i)}\,
h[\phi(0)]\,\Big\rangle,
\end{eqnarray}
where the state $\ket{\phi}$ is given in the form
$\ket{\phi}=\phi(z=0)\ket{0}$ for an operator $\phi(z)$, and $h[\cdots]$
denotes the conformal mapping corresponding to the function
$u=h(z)=2z/(1-z^2)$, which 
maps the unit disk $\abs{z}\leq 1$ into the whole complex $u$-plane.
The correlation function of the right hand-side is defined in the
$u$-plane and the momenta $k_L$ and $k_R$ are given by
$k_L=(m/R+wR/\alpha')/2$ and $k_R=(m/R-wR/\alpha')/2$.

From the definition (\ref{Eq:mwRdef}), we can write the state
$\bra{m,w;R}$ by using oscillator expression:
\begin{eqnarray}
\label{Eq:clXstate}
 \bra{m,w;R}=N\,\bra{-\frac{m}{R}}\exp
\sum_{n=1}^\infty (-1)^n\left\{
-\frac{1}{2n}\alpha_n \alpha_n
-\frac{\sqrt{2\alpha'}}{n}\frac{m}{R}\alpha_{2n}
-\frac{2i\sqrt{2\alpha'}}{2n-1}\frac{wR}{\alpha'}
\alpha_{2n-1}\right\},
\end{eqnarray}
where $N$ denotes a normalization factor depending on $m$, $w$ and
$R$, and the state $\bra{-m/R}$ denotes the eigenstate of $\hat{p}$, 
that is  $\bra{-m/R} \hat{p} =-m/R \bra{-m/R}$. The derivation
is put in Appendix A. 
From the oscillator expression, we observe that the momentum is
conserved in this open-closed string coupling but the winding number is
not because the closed string state 
contains only the zero mode eigenstate $\bra{-m/R}$. We can obtain a
closed string coupling to a Dirichlet open string by the $T$-dual
transformation, $m\leftrightarrow w$ and $R\leftrightarrow \alpha'/R'$.
This closed string state is represented by
\begin{eqnarray}
 \bra{w,m;\frac{\alpha'}{R'}}=\bra{-\frac{wR'}{\alpha'}}\exp
\sum_{n=1}^\infty (-1)^n\left\{
-\frac{1}{2n}\alpha_n \alpha_n
-\frac{\sqrt{2\alpha'}}{n}\frac{wR'}{\alpha'}\alpha_{2n}
-\frac{2i\sqrt{2\alpha'}}{2n-1}\frac{m}{R'}
\alpha_{2n-1}\right\}.
\end{eqnarray}
In this $T$-dual picture, the winding number is conserved but the
momentum is not. This is a consistent result with the coupling of a
closed string in a bulk and an open string on a D-brane.

Now, let us consider open-closed string couplings
at the classical background corresponding to the solution (\ref{Eq:sol}):
\begin{eqnarray}
\label{Eq:VPsigauge}
\bra{V(m,w;R)}\ket{\Psi}=
\bra{V(m,w;R)}\ket{\Psi_0(\lambda)}
+\bra{V(m,w;R)}e^{B(\lambda)}\ket{\Psi'}.
\end{eqnarray}
Note that the action of $\Psi'$ has the original form with the ordinary
BRS charge as discussed in the previous section.
First, we can show that the inhomogeneous term
$\bra{V(m,w;R)}\ket{\Psi_0(\lambda)}$ vanishes in the following. 
Differentiating the solution $\Psi_0(\lambda)$
with respect to $\lambda$, we find
\begin{eqnarray}
\label{Eq:dPsi}
\frac{d}{d\lambda}\ket{\Psi_0(\lambda)}
&=&-\Q'(\lambda)\,
\frac{i}{2\sqrt{\alpha'}}X_L(F)\ket{I},
\end{eqnarray}
where we have used commutation relations of
$V_L$, $C_L$ and $X_L$. Moreover, using the oscillator expression of the
closed string state, we can find the modified BRS invariance
of $\bra{V(m,w;R)}$:
\begin{eqnarray}
\label{Eq:VQB}
 \bra{V(m,w;R)}\Q'(\lambda)=0.
\end{eqnarray}
The proof is given in Appendix B. 
From (\ref{Eq:dPsi}) and (\ref{Eq:VQB}), we obtain a similar equation to
that of the vacuum energy (\ref{Eq:dS}):
\begin{eqnarray}
 \frac{d}{d\lambda}\bra{V(m,w;R)}\ket{\Psi_0(\lambda)}=0.
\end{eqnarray}
Similarly, we verify that
the term $\bra{V(m,w;R)}\ket{\Psi_0(\lambda)}$
is to be zero irrelevant to $\lambda$.
Here, it should be noticed that this evaluation
is rather formal as well as that of the vacuum energy. From
direct calculation using oscillator expression, we find that this term
also is an indefinite quantity. This question remains open.

As a result, at the classical background, the closed string state is
changed by the transformation generated by $B(\lambda)$ from the
original form (\ref{Eq:VmwR}).
Since the operator $B(\lambda)$ contains oscillators of only the
compactified direction, we have only to calculate the 
effect of $B(\lambda)$ on $\bra{m,n;R}$ in order to evaluate the
right hand-side of (\ref{Eq:VPsigauge}).
For the function $F(z)=(z+1/z)/z$, using the oscillator expression
(\ref{Eq:Blamd}) and (\ref{Eq:clXstate}), we can easily find that
\begin{eqnarray}
\label{Eq:mwRB}
 \bra{m,w;R}B(\lambda)= i\frac{wR}{\alpha'}\,2\sqrt{\alpha'}\lambda
\,\bra{m,w;R}.
\end{eqnarray}
Then, we obtain a final expression of the open-closed string coupling:
\begin{eqnarray}
\label{Eq:VPsiphase}
 \bra{V(m,w;R)}\ket{\Psi}
=\exp\left(
{i\frac{wR}{\alpha'}\,2\sqrt{\alpha'}\lambda}
\right)
 \bra{V(m,w;R)}\ket{\Psi'}.
\end{eqnarray}
Thus, the classical solution for the marginal deformation
causes the phase shift which is proportional to the winding number
in the open-closed strings coupling. This
phase shift (\ref{Eq:VPsiphase}) completely agrees 
with the effect of the marginal deformation 
of the Wilson line flux \cite{rf:RS}. 
In the T-dual picture, the coupling is rewritten as
 \begin{eqnarray}
\label{Eq:VPsiphaseT}
 \bra{V(w,m;\alpha'/R')}\ket{\Psi}
=\exp\left(
{i\frac{m}{R'}\,2\sqrt{\alpha'}\lambda}
\right)
 \bra{V(w,m;\alpha'/R')}\ket{\Psi'}.
\end{eqnarray}
This implies that the marginal deformation moves the position of
D-brane as usual \cite{rf:RS}.

Now consider how the closed string coupling
at the critical radius  is changed by string field expansion around
$\Psi_0^2$ and the 
string field redefinition (\ref{Eq:Blamdcr}). In this case too, we find
that the inhomogeneous term becomes zero for any $\lambda$. Then, we
have only to consider the effect of the string field redefinition on the
closed string state.

For simplicity, we consider the case that the closed string
state is given by $\bra{V}=\bra{V(0,1)}-\bra{V(0,-1)}$, where we have
used 
$\alpha'=1$ 
unit and omitted the radius $R=\sqrt{\alpha'}$ in the expression.
From the definition (\ref{Eq:mwRdef}) ,
the compactified sector of $\bra{V(0,1)}$ is defined by a
midpoint insertion of the vertex operator
\begin{eqnarray}
{\cal V}(u,\bar{u})=e^{\frac{i}{2}X(u)-\frac{i}{2}X(\bar{u})}.
\end{eqnarray}
The effect of $\gamma_0$ on $\bra{V(0,1)}$ can be evaluated by
deforming the integration path of $\gamma_0$ to contours around $\pm i$
in the u plane and calculating OPE:
\begin{eqnarray}
-\left(
\oint_{C_i}\frac{du}{2\pi i}\,J^3(u)
+\oint_{C_{-i}}\frac{du}{2\pi i}\,J^3(u)\right)\,{\cal V}(i,-i)
=-\frac{i}{2}e^{-\frac{i}{2}X(i)-\frac{i}{2}X(-i)}
+\frac{i}{2}e^{\frac{i}{2}X(i)+\frac{i}{2}X(-i)},
\end{eqnarray}
and then we find
\begin{eqnarray}
 \bra{V(0,1)}\gamma_0=\frac{i}{2}\bra{V(1,0)}-\frac{i}{2}\bra{V(-1,0)}.
\end{eqnarray}
From this and similar equations, we obtain
\begin{eqnarray}
\label{Eq:Vgamma}
 \bra{V(0,\pm 1)}e^{i\theta\gamma_0}&=&
  \cos^2 \frac{\theta}{2} \bra{V(0,\pm 1)}
  -\sin^2 \frac{\theta}{2} \bra{V(0,\mp 1)}\nn
&&  -\frac{1}{2}\sin \theta\left\{
   \bra{V(1,0)}-\bra{V(-1,0)}\right\}.
\end{eqnarray}
As a result, we find that the state
$\bra{V}=\bra{V(0,1)}-\bra{V(0,-1)}$ is invariant under
the rotation of $\gamma_0$. Furthermore,
it can be easily seen that $\bra{V}$ is a singlet state of
the $SU(2)$ symmetry generated by $\alpha_0$, $\beta_0$ and $\gamma_0$.
Hence, this closed string state $\bra{V}$ is regarded as a superposition
of states with the winding number $\pm 1$ or with the momentum $\pm 1$
in the T-dual picture, and this interpretation is
unchanged under the $SU(2)$ transformation.

We can rewrite $B'(\lambda)$ as a $SU(2)$ rotation of $B(\lambda)$
by using the commutation relations (\ref{Eq:comrel}):
\begin{eqnarray}
 B'(\lambda)= e^{-i\frac{\pi}{2}\gamma_0}\,B(\lambda)\,
e^{i\frac{\pi}{2}\gamma_0}.
\end{eqnarray}
Then, the closed string state
is transformed by the string field redefinition (\ref{Eq:Blamdcr}) to 
\begin{eqnarray}
 \bra{V}e^{B'(\lambda)}&=& 
 \bra{V}e^{B(\lambda)}\,e^{i\frac{\pi}{2}\gamma_0}\nn
 &=& e^{i\,2\lambda}\bra{V(0,1)}\,e^{i\frac{\pi}{2}\gamma_0}
    -e^{-i\,2\lambda}\bra{V(0,-1)}\,e^{i\frac{\pi}{2}\gamma_0},
\end{eqnarray}
where we have used the $SU(2)$ invariance of $\bra{V}$
and the transformation law (\ref{Eq:mwRB}).
Applying (\ref{Eq:Vgamma}), we find final
expression of the transformation law of the closed string state:
\begin{eqnarray}
\label{Eq:VB}
 \bra{V}e^{B'(\lambda)} = \cos 2\lambda \bra{V}
 -i\sin 2\lambda \bra{V'},\ \ \ 
 \bra{V'}= \bra{V(1,0)}-\bra{V(-1,0)},
\end{eqnarray}
where $\bra{V'}$ is a superposition of states with the
momentum $\pm 1$ or with the winding number $\pm 1$ in the T-dual
picture.

Now that the closed string state at the classical background is given,
let us consider the interpretation of the resulting open-closed 
string coupling. Supposed that the closed string state has
momentum but no winding number in the coupling,
and then the arguments
$\pm 1$ of both of $\bra{V}$ and $\bra{V'}$ are to be momentum in
the compactified direction. 
Then, the coupling $\bra{V}\ket{\Psi}$
can be regarded as that of the closed string state
to the open string field on an array of D-branes with Dirichlet boundary
condition on $X$, and
$\bra{V'}\ket{\Psi}$ corresponds to a closed string coupling to the open
string field with Neumann boundary condition on $X$.
As a result, the transformation law (\ref{Eq:VB}) implies that
string field condensation on an array of D-branes causes annihilation
of the D-brane array at $\lambda=\pi/4$ and creation of the same
D-brane array at $\lambda=\pi/2$. This effect of string field
condensation is the same as that of the marginal deformation associated
with the current $J^2$ \cite{rf:Sen}.

Finally, let us reexamine the relation between the marginal deformation
parameter in conformal field theory and the corresponding
parameter $\lambda$ in string field theory.
As in the previous section, the string field condensation
associated with
$J^1$ produces the momentum shift as
$p \rightarrow p+(\lambda_i-\lambda_j)/\pi\sqrt{\alpha'}$ if we
introduce the Chan-Paton indices. Comparing this effect with that of
conformal field theory, we conclude that string field condensation
of (\ref{Eq:sftcond}) corresponds to adding to the world-sheet action
the boundary term
\begin{eqnarray}
\label{Eq:CFTbt}
 i\,\frac{\lambda}{\pi\sqrt{\alpha'}}
\int dt \,\partial_t X(t)=
i\frac{\theta}{2\pi R}\int dt \,\partial_t X(t)
=
i\frac{\theta R'}{2\pi \alpha'}\int dt \,\partial_t X(t),
\end{eqnarray}
where we introduce the parameter $\theta=2\lambda R/\sqrt{\alpha'}
=2\sqrt{\alpha'} \lambda/R' $.
From the transformation law (\ref{Eq:VPsiphaseT}) in the
T-dual picture, it follows that this
string field condensation induces the shift of D-brane position as
$x\rightarrow x+2\sqrt{\alpha' \lambda}=x+\theta R'$.
This result completely agrees with the effect of the boundary term
(\ref{Eq:CFTbt}) in conformal field theory.

At the critical radius,\footnote{We have used $\alpha'=1$ unit.}
the boundary term (\ref{Eq:CFTbt}) can be
transformed to
\begin{eqnarray}
  \frac{2\lambda}{\pi}
\int dt \,\cos X(t),
\end{eqnarray}
by a $SU(2)$ rotation. This boundary term creates or annihilate an array
of D-branes at $2\lambda/\pi=1/2$ \cite{rf:CKLMPT,rf:RS,rf:Sen}. This
effect has also complete 
agreement with the result obtained from (\ref{Eq:VB}) in string field
theory. 

\section{Discussions}

We studied the analytic classical solutions in string field theory which
corresponds to boundary marginal deformations in conformal field
theory. In particular, we evaluated the closed string couplings to the
open string field in the classical backgrounds and then we confirmed
that the resulting couplings have the periodic properties expected from
conformal field theory. 

In the open-closed string coupling, we did not consider the vertex
operator with derivatives of the compactified direction. However, we 
should calculate more generic closed string vertex operators
in order to prove the correspondence between the classical
solution and the marginal deformation.
At the critical radius, we should evaluate
the open-closed string coupling in which the closed string state
associates to a more general representation of the $SU(2)$ group. 
These problems remain to be resolved.

These analytic classical solutions have the manifest relation to the
marginal deformations as far as string field redefinitions and
open-closed string couplings are concerned. To make the relation
more precise, we must understand the vacuum energy of the classical
solutions, whereas we can not evaluate it directly because of its
indefiniteness. However, there is one possible method to speculate
it. We can determine the vacuum energy indirectly by 
investigating the tachyon vacuum in the theory expanding string
field around the classical solution. Since the Wilson lines flux does
not affect the D-brane tension, the vacuum energy of the
tachyon vacuum could be independent of the marginal parameter of the
solution. We could confirm this conjecture by using the level truncation 
analysis as discussed in ref.~\cite{rf:tomo}.

In the level truncated theory, the graviton coupling directly to the
D-brane has been obtained from the energy momentum tensor in string
field theory \cite{rf:SZmarginal}.
This graviton coupling can not be calculated 
by considering gauge invariant operators as discussed in this paper. 
For the analytic solutions, it has been still unclear how the direct
closed string coupling to the D-brane can be derived as pointed out in
ref.~\cite{rf:SZmarginal}.

\section*{Acknowledgements}
The authors would like to thank H.~Hata, Y.~Igarashi, K.~Itoh and
M.~Kenmoku 
for useful discussions. They thank also the Yukawa Institute for
Theoretical Physics at Kyoto University. Discussions during the YITP
workshop YITP-W-04-03 on ``Quantum Field Theory 2004" were useful to
complete this work.  

\appendix

\section{Oscillator Expression of
Closed String States}

We will derive the oscillator expression (\ref{Eq:clXstate}) of the state
$\bra{m,w;R}$. From the definition (\ref{Eq:mwRdef}), we can represent
this state as
\begin{eqnarray}
\label{Eq:mwRosc}
 \bra{m,w;R}=N\bra{\frac{-m}{R}}\exp\left(
\frac{1}{2}\sum_{n,l=1}^\infty \bar{N}_{nl}\alpha_n \alpha_l
+\frac{m}{R} \sum_{n=1}^\infty \bar{N}_n \alpha_n
+\frac{wR}{\alpha'} \sum_{n=1}^\infty \bar{D}_n \alpha_n \right),
\end{eqnarray}
where $N$ denotes a normalization factor depending on $m$,$w$ and $R$
, and $\bar{N}_{nl}$, $\bar{N}_n$ and $\bar{D}_n$ are constants.

Considering the inner product between the state (\ref{Eq:mwRosc}) and
$\ket{k/R}$ and  mapping it to the $u$ plane, we can find
the equation 
\begin{eqnarray}
\label{Eq:N}
 N \left<\frac{-m}{R} \right| 
   \left.\frac{k}{R} \right>=
\left(\partial h(0)\right)^{\alpha'\frac{k^2}{R^2}}
\left<\,e^{ik_L X(i)+ik_R X(-i)}\,
e^{i \frac{k}{R} X(0)}\,\right>.
\end{eqnarray}
Then, we can obtain the normalization factor $N$ by calculating the
correlation function of the right hand-side.

We can calculate $\bar{N}_n$ and $\bar{D}_n$ by multiplying
$\alpha_{-n}\ket{k/R}$ to (\ref{Eq:mwRosc}):
\begin{eqnarray}
\label{Eq:preND}
&&
 n\left(\frac{m}{R}\bar{N}_n
 +\frac{wR}{\alpha'}\bar{D}_n\right) N \left<\frac{-m}{R} \right| 
   \left.\frac{k}{R} \right> \nn
&=&\left(\partial h(0)\right)^{\alpha'\frac{k^2}{R^2}}
\frac{i}{\sqrt{2\alpha'}} \oint_{C_0}\frac{dz}{2\pi i}
z^{-n}\,\partial h(z) 
\left<\,e^{ik_L X(i)+ik_R X(-i)}\,\partial X(u)\,
e^{i \frac{k}{R} X(0)}\,\right>.
\end{eqnarray}
The correlation function in the integrand can be evaluated as
\begin{eqnarray}
\label{Eq:correlfn}
&&
 \left<\,e^{ik_L X(i)+ik_R X(-i)}\,\partial X(u)
e^{i \frac{k}{R} X(0)}\,\right> \nn
&=& \left\{
i 2\alpha'\frac{m}{R}\frac{1}{u(1+u^2)}
+2\alpha' \frac{wR}{\alpha'}\frac{1}{1+u^2}\right\}
 \left<\,e^{ik_L X(i)+ik_R X(-i)}\,
e^{i \frac{k}{R} X(0)}\,\right>,
\end{eqnarray}
where we have used the momentum conservation $m+k=0$. 
Substituting (\ref{Eq:correlfn}) into (\ref{Eq:preND}) and using 
(\ref{Eq:N}), we find the expression for $\bar{N}_n$ and $\bar{D}_n$ as
\begin{eqnarray}
 \bar{N}_n
&=&-\frac{\sqrt{2\alpha'}}{n}\oint_{C_0}\frac{dz}{2\pi i}
z^{-n}\,\partial h(z)\,\frac{1}{u(1+u^2)}, \\
 \bar{D}_n
&=&i \frac{\sqrt{2\alpha'}}{n}\oint_{C_0}\frac{dz}{2\pi i}
z^{-n}\,\partial h(z)\,\frac{1}{u(1+u^2)}.
\end{eqnarray}
Since the mapping $u=h(z)$ is given by $h(z)=2z/(1-z^2)$, we can expand
the integrands as
\begin{eqnarray}
 z^{-n}\,\partial h(z)\,\frac{1}{u(1+u^2)} 
   = z^{-n-1} \frac{1-z^2}{1+z^2}
   = z^{-n-1}\left(1+2\sum_{n=1}^\infty (-1)^n z^n
    \right),\nn
 z^{-n}\,\partial h(z)\,\frac{1}{1+u^2}=z^{-n}\frac{2}{1+z^2}
  =z^{-n-1}\left(-2\sum_{n=1}^\infty (-1)^n z^{2n-1}\right).
\end{eqnarray}
Then, we can obtained the final expression for $\bar{N}_n$ and
$\bar{D}_n$: 
\begin{eqnarray}
 \bar{N}_{2n}=-\frac{\sqrt{2\alpha'}}{n},\ \ \ 
 \bar{D}_{2n-1}=-\frac{2i\sqrt{2\alpha'}}{2n-1},\ \ \ 
 \bar{N}_{2n-1}=0,\ \ \ \bar{D}_{2n}=0\ \ \ 
 (n=1,2,3,\cdots).
\end{eqnarray}

For $\bar{N}_{nl}$, by multiplying the state
$\alpha_{-n}\alpha_{-l} \ket{k/R}$ to (\ref{Eq:mwRosc}),
we find the equation 
\begin{eqnarray}
&&
 n\,l\,\left\{
\bar{N}_{nl}+\left(\frac{m}{R}\bar{N}_n+\frac{wR}{\alpha'}\bar{D}_n
\right)\left(\frac{m}{R}\bar{N}_l+\frac{wR}{\alpha'}\bar{D}_l\right)
\right\}\,
 N \left<\frac{-m}{R} \right| 
   \left.\frac{k}{R} \right> \nn
&=&
\left(\partial h(0)\right)^{\alpha'\frac{k^2}{R^2}}
\left(\frac{i}{\sqrt{2\alpha'}}\right)^2
\oint_{C_0}\frac{dz}{2\pi i}
\,\oint_{C_0}\frac{d\tilde{z}}{2\pi i}
\,z^n\,\partial h(z)\,\,{\tilde{z}}^l
\,\partial h(\tilde{z})
\times \nn
&&\times \left<\,e^{ik_L X(i)+ik_R X(-i)}\,
\partial X(u)\,\partial X(\tilde{z})\,
e^{i \frac{k}{R} X(0)}\,\right>.
\end{eqnarray}
Calculating the correlation function, we can find the expression
\begin{eqnarray}
\bar{N}_{nl}=\frac{1}{nl}
\oint_{C_0}\frac{dz}{2\pi i}
\,\oint_{C_0}\frac{d\tilde{z}}{2\pi i}
\,z^n\,\partial h(z)\,\,{\tilde{z}}^l
\,\partial h(\tilde{z})\frac{1}{(u-\tilde{u})^2}.
\end{eqnarray}
Similarly, by using the expansion
\begin{eqnarray}
 \partial h(z)\,\partial h(\tilde{z})\,
\frac{1}{(u-\tilde{u})^2}=\frac{1}{(z-\tilde{z})^2}
+\frac{1}{(1+z\tilde{z})^2}
=\frac{1}{(z-\tilde{z})^2}
-\sum_{n=1}^\infty n (z\tilde{z})^{n-1},
\end{eqnarray}
we obtain the final expression
for $\bar{N}_{nl}$
\begin{eqnarray}
 \bar{N}_{nl}=-\frac{(-1)^n}{n}\delta_{n,l}.
\end{eqnarray}

\section{Modified BRS Invariance of
Closed String States}

We will prove that
the modified BRS charge (\ref{Eq:newBRS}) annihilates
the state (\ref{Eq:VmwR}).

The function $F(z)$ satisfying $F(-1/z)=z^2 F(z)$ can be
expanded as \cite{rf:TT2}
\begin{eqnarray}
 F(z)=\sum_{n=1}^\infty f_n (z^n-(-1/z)^n)z^{-1}.
\end{eqnarray}
Since the function $G(z)=F(z)^2$ satisfies $G(-1/z)=z^4 G(z)$, we can
expand $F(z)^2$ as
\begin{eqnarray}
 F(z)^2=\sum_{n=0} g_n (z^n+(-1/z)^n)z^{-2}.
\end{eqnarray}
Substituting these expansions into (\ref{Eq:newBRS}), 
we can write the modified BRS charge 
by using oscillator expression:
\begin{eqnarray}
 \Q'(\lambda)=\Q-\lambda \sum_{n=1}^\infty f_n\,(t_n-(-1)^n t_{-n})
+\frac{\lambda^2}{4}\sum_{n=0}^\infty g_n\,(c_n+(-1)^n c_{-n}),
\end{eqnarray}
where $t_n$ denote oscillators of the current $J(z)$ and they are
written as
\begin{eqnarray}
 t_n=\frac{1}{\sqrt{2}}\sum_{m=-\infty}^\infty c_{n+m}\alpha_{-m}.
\end{eqnarray}

By definition, it can be easily seen that $\bra{V(m,w;R)}$ is a BRS
invariant state. Then, in order to see the modified BRS invariance
(\ref{Eq:VQB}), we have only to 
prove the following equations
\begin{eqnarray}
\label{Eq:Vghcn}
&&
\bra{V_{gh}}(c_n+(-1)^n c_{-n})=0, \\
\label{Eq:Vtn}
&&
\bra{m,w;R}\otimes \bra{V_{gh}}(t_n-(-1)^n t_{-n})=0.
\end{eqnarray}
The first equation (\ref{Eq:Vghcn}) can be seen by using
oscillator expression  of $\bra{V_{gh}}$ \cite{rf:HI,rf:TZ2}:
\begin{eqnarray}
 \bra{V_{gh}}=
\bra{0}c_{-1}c_0\exp \sum_{n=1}^\infty
(-1)^{n+1} c_n b_n.
\end{eqnarray}
From the expression (\ref{Eq:clXstate}), it follows that
\begin{eqnarray}
\label{Eq:mwRae}
 \bra{m,w;R}(\alpha_{2n}+\alpha_{-2n})&=&
-(-1)^n\,2\sqrt{2\alpha'}\,\frac{m}{R}\,\bra{m,w;R},\\
\label{Eq:mwRao}
 \bra{m,w;R}(\alpha_{2n-1}-\alpha_{-2n+1})&=&
(-1)^n\,2i \sqrt{2\alpha'}\,\frac{wR}{\alpha'}
\,\bra{m,w;R}.
\end{eqnarray}
Using (\ref{Eq:Vghcn}), (\ref{Eq:mwRae}) and (\ref{Eq:mwRao}), we can
calculate the left hand-side of (\ref{Eq:Vtn}) and find that it becomes
zero. Hence, we can see that the equation (\ref{Eq:VQB}) holds.


\end{document}